\documentclass[aps, showpacs]{revtex4}

\begin{document}
\title{On local duality invariance in electromagnetism}
\author{S. C. Tiwari \\
Institute of Natural Philosophy \\
1 Kusum Kutir, Mahamanapuri \\
Varanasi 221005, India }
\begin{abstract}
Duality is one of the oldest known symmetries of Maxwell equations. In recent years the significance of duality symmetry has been recognized in superstrings and high energy physics and there has been a renewed interest on the question of local duality rotation invariance. In the present paper we re-visit global duality symmetry in the Maxwell action and delineate the ambiguous role of gauge invariance and time locality. We have recently demonstrated that local duality invariance in a Lorentz covariant form can be carried out in the Maxwell equations. In this paper it is shown that in the four-pseudo vector Lagrangian theory of Sudbery a local duality generalization can be naturally and unambiguously implemented and the Euler-Lagrange equations of motion are consistent with the generalized Maxwell field equations. It is pointed out that the extension of Noether theorem in full genrality for a vector action is an important open problem in mathematical physics. Physical consequences of this theory for polarized light and topological insulators are also discussed.
 
\end{abstract}
\pacs{11.30.-j, 03.50.De}
\maketitle
\section{\bf Introduction}
Electric-magnetic duality in supersymmetric gauge theories and the profound role of duality in the second superstring revolution in 1990s generated a lot of interest amongst theoretical high energy physicists in this subject; see a nice review with extensive citation to the literature by Duff \cite{1}. Witten articulated the conceptual development of duality paradigm with emphasis on superstring theory in a lucid expository article in \cite{2}. Interestingly Zee in his book \cite{3} remarks that, 'In contrast, according to one of my distinguished condensed matter colleagues, the important notion of duality is still underappreciated in the condensed matter physics community'. However SL(2, Z) symmetry has been discussed in connection with topological aspects of Hall effect phase transitions by Fradkin and Kivelson \cite{4}. Fradkin (in private communication) incidentally points out that Zee's remarks are not correct, and the condensed matter physicists had long ago realized the significance of duality. Our view is that duality has wider significance in theoretical physics. Note that duality symmetry prior to these developments was primarily associated with electromagnetism and optical phenomena \cite{5, 6}. Recently Bunster and Henneaux \cite{7} raised the question whether the electric-magnetic duality could be gauged and concluded that it could not be. However nearly two decades ago \cite{8} a local duality invariant formulation was presented generalizing four vector action integral proposed by Sudbery \cite{9}. Not only this, local duality gauge theory in the Schroedinger-like form of Maxwell equations following \cite{10} was discussed in \cite{11} to shed light on Pancharatnam topological phase in optics \cite{12}.

Deser has further demonstrated \cite{13} the no-go result of \cite{7} in the canonical formulation \cite{14} of the Maxwell theory of electromagnetism. We do not dispute this conclusion of \cite{7, 13} but emphasize that local duality gauge theory is not unphysical and could be implemented both at the level of equations of motion and manifestly in the action \cite{8}. The aim of the present paper is to resolve this seemingly paradoxical situation and offer new insights on the question of local duality invariance. The key issue in this context would be to delineate the fundamental principles involved and the independent field variables assumed. Though we are only concerned with duality in electromagnetism in the light of its wider ramifications first we give a brief description of this idea in next section as it is understood in diverse fields. Global duality invariance in which the rotation angle is a constant parameter is reviewed in Sec. III. In Sec. IV the variational principle of Anderson and Arthurs \cite{15} and Rosen \cite{16} is briefly discussed that motivated the work in \cite{10}. Salient features in Sudbery's theory \cite{10} and the generalized local gauge invariant action are presented next. New field equations in covariant form are derived and their consistency with local duality invariant Maxwell equations given recently \cite{17} is shown. In Sec. V physical significance of duality gauge theory is discussed and possible applications are outlined with the concluding remarks.

\section{\bf Preliminaries on duality symmetry}
If symmetry symbolizes beauty Maxwell equations are most beautiful. These are invariant under a 15-parameter group of conformal transformations with 10-parameter Lorentz sub-group \cite{18}. Remarkably both general covariance and metric-free topological invariance co-exist in the Maxwell equations. Regarding electric-magnetic duality Deser \cite{13} rightly remarks that it is an 'ancient lore'. Lipkin discovered ten new conserved quantities for vacuum Maxwell equations \cite{19} and named them zilch of electromagnetic field. Note that conservation laws are intimately related with symmetry. Of course, the most fruitful symmetry that paved the path for modern unified theories is that of gauge invariance once the electromagnetic potential is introduced \cite{20}.

The complete set of vacuum Maxwell equations with sources has two manifest asymmetries: absence of magnetic source terms and sign asymmetry in the time derivative. In 1931 Dirac postulated magnetic monopole with the main aim to explain electric charge quantization. Later in a comprehensive theory in 1948 \cite{21} the role of symmetry was underlined by him. Insightful remarks in the last section of this paper deserve attention. A question is raised whether an elementary particle could possess both electric and magnetic charges, and the answer is left undecided in the absence of a theory that accounts for self interactions. He further notes that the theory is 'essentially symmetrical between electric charges and magnetic poles', however there is a difference due to the different values of the coupling strengths $e^2/\hbar c = 1/137$ and $g^2/\hbar c = 137/4$. We reproduce two statements relevant to duality from this article: 'However, one could work equally well with the roles of charges and poles interchanged'. And, 'The final result would be an equivalent quantum electrodynamics referred to a different representation'. As explained below Dirac could be credited anticipating the most modern version of duality principle. In the passing we just mention a discussion on time asymmetry in \cite{22} and focus our attention on the notion of duality in the following.

No experiment till date has found any evidence for the existence of a magnetic monopole, yet monopole continues to be a linking thread in the evolution of duality principle. First, a monopole solution was discovered in non-abelian classical Yang-Mills theory, and then in unified gauge theories monopole with a huge mass of the order of $10^{19} ~Gev ~c^{-2}$ was predicted. Recall that Schwinger's speculation on particles having both electric and magnetic charges named dyons by him by implication could be utilized to understand absence of magnetic charges and currents making duality rotation \cite{23}. A class of gauge theories admit stable particles with electric charges $Q_e = p e$ and magnetic charges $Q_m= qg$ where p, q are integers, with mass formula of the form $M=\sqrt {Q_e^2+Q_m^2}$. In 1977 Montonen and Olive interpreted this in terms of a symmetry between the exchange of electric to magnetic charges and vice versa. In such a symmetry the coupling constant transforms as $\alpha \leftrightarrow \alpha ^{-1}$ and exchanges elementary quanta with collective excitations. For weak coupling $\alpha$ electric charge is elementary and monopole is a soliton-like excitation whereas for strong coupling monopole is elementary and electric charge emerges as a collective excitation. In four dimensions this duality can be realized only in supersymmetric gauge theories. In a stronger version one has the notion of self-duality: the dual theory is the same as the original one. We refer to Duff \cite{1} for a discussion on $N=4$ supersymmetric self-dual gauge theory. In a supersymmetric gauge theory there is an additional parameter, namely the vacuum angle $\theta$ that can be combined with charge $e$ to define a complex parameter $S$
\begin{equation}
 S= \frac{\theta}{2\pi} + i \frac{4\pi}{e^2}
\end{equation}
where magnetic charge $Q_m =\frac{n}{e}$ and electric charge $Q_e =e(m+\frac{n\theta}{2\pi})$. Here n, m are integers.
Electric-magnetic duality forms a group of order 2 and $Z$ in SL(2, Z) signifies that the matrix elements in $2 \times 2$ matrices $\left(\begin{array}{cc} a & b \\ c & d\\
                     
                    \end{array}\right)$
that form the modular group are integers. Note that the determinant of the matrices is one, that is $ad - bc =1$. Townsend \cite{24} in a pleasant popular-level but insightful article on duality remarks regarding SL(2, Z) symmetry: {\it By an abuse of language it has become customary to refer also to this generalization of electromagnetic duality (and others) as a 'duality'.}

The action of SL(2, Z) on parameter $S$ is
\begin{equation}
 S \rightarrow \frac{aS+b}{cS + d}
\end{equation}
In superstring theory analogue of complex coupling constant is a complex scalar field in which angle $\theta$ corresponds to the vacuum expectation value of axion field and charge to that of a dilaton field. S-duality in superstrings becomes a transformation law for axion-dilaton field. The picture that seems to emerge is that there is an underlying unity amongst various superstrings in what is called M theory. Last sentences in \cite{1, 24} aptly capture the essence when Duff sees the role of duality in 'unification via diversification' and Townsend finds it ironic that 'newly emerging unity is the result of a renewed interest in the old idea of duality'. Unfortunately the envisaged dream of unification remains elusive so far.

\section{\bf Duality invariance of Maxwell field equations and action} 

Originally electric-magnetic duality meant ${\bf E} \rightarrow {\bf B}$ and ${\bf B} \rightarrow -{\bf E}$. It can be extended to a more general rotation with arbitrary constant rotation angle $\zeta$ such that
\begin{equation}
 {\bf E} \rightarrow {\bf E}~ cos \zeta + {\bf B} ~sin\zeta
\end{equation}
\begin{equation}
 {\bf B} \rightarrow - {\bf E} ~sin \zeta + {\bf B} ~cos \zeta
\end{equation} 
It can be easily verified that source-free vacuum Maxwell equations
\begin{equation}
 {\bf \nabla.E} = 0
\end{equation}
\begin{equation}
 {\bf \nabla} \times {\bf B} - \frac {\partial {\bf E}}{\partial t} =0
\end{equation}
\begin{equation}
{\bf \nabla} \times {\bf E} + \frac {\partial {\bf B}}{\partial t} =0 
\end{equation}
\begin{equation}
 {\bf \nabla.B} = 0
\end{equation}

are invariant under this transformation. However, the action
\begin{equation}
 I = \frac{1}{2} \int ({\bf E}^2 - {\bf B}^2)~ d^4 x
\end{equation}
is not manifestly invariant under duality rotation. This would seem puzzling, however the authors in \cite{7} emphasize that it is erroneous to say that duality is not a symmetry of action but only that of equations of motion. There is a technical subtelity involved in this assertion, therefore this section is devoted to revisit this issue with a fresh outlook and also to bring to the notice significant past contributions that gave deep physical insights.

Let us consider infinitesimal rotation $\delta \zeta$ that reduces (3) and (4) to
\begin{equation}
 {\bf E} \rightarrow {\bf E} + {\bf B} ~\delta\zeta
\end{equation}
\begin{equation}
 {\bf B} \rightarrow - {\bf E} ~\delta\zeta + {\bf B}
\end{equation} 
then action (9) changes by
\begin{equation}
 \delta I = 2 \delta \zeta \int {\bf E.B}~ d^4x 
\end{equation}
neglecting higher order terms in $\delta \zeta$. As it stands, expression (12) shows that action is not invariant even under infinitesimal duality rotation. Deser and Teitelboim \cite{25} point out that ${\bf E.B}$ term could be re-written as $\partial_\mu ~C^\mu$ where
\begin{equation}
 C^\mu = \epsilon^{\mu\nu\rho\sigma} A_\nu \partial_\rho A_\sigma
\end{equation}
since according to them '$F_{\mu\nu}$ is only a shorthand for $\partial_\mu A_\nu -\partial_\nu A_\mu$'. Here $\epsilon ^{\mu\nu\rho\sigma}$ is Levi-Civita tensor in four dimension and $F^{\mu\nu}$ is electromagnetic field tensor. It then follows that $\delta I$ vanishes and action is invariant. In the case of finite rotation though the ${\bf E.B}$ term would again arise and could be made to vanish for the identical reason the action transforms to
\begin{equation}
 I \rightarrow (cos^2 \zeta -sin^2 \zeta) ~I
\end{equation}
and clearly not invariant.

Invariance of action (9) in both second-order Lagrangian and first-order Hamiltonian forms is proved in \cite{25}, and off-shell duality invariance is further clarified in \cite{14}. In two-vector potential formalism the action (9) assumes a manifestly duality invariant form, see Sec. IIA of \cite{7}
\begin{equation}
 I_V =\frac{1}{2} \int (\epsilon _{ab} {\bf B}^a . \dot{{\bf A}}^b - \delta_{ab} {\bf B}^a. {\bf B}^b) ~ d^4x
\end{equation}
Note that Kronecker symbol $\delta_{ab}$ and Levi-Civita tensor $\epsilon_{ab}$ are invariant under rotation in two dimension and $a, b =1, 2$. Here over-dot denotes time derivative and
\begin{equation}
 {\bf B}^a = {\bf\nabla} \times {\bf A}^a
\end{equation}
with ${\bf B}^1$ and $-{\bf B}^2$ identified as magnetic and electric fields respectively and the duality rotation corresponds to the rotation in ${\bf A}^1, {\bf A}^2$.

The main argument to implement duality transformation in action is that one has to consider the basic dynamical field variables, and satisfy time locality requirement \cite{25}. In the action integral $A_\mu$ is a fundamental dynamical variable for Lagrangian form and $A_\mu$ and its canonically conjugate variable for the Hamiltonian formulation. At this point it would be of interest to briefly mention other significant contributions. Calkin using Noether's theorem arrived at an important result in \cite{5}. He asked the question: What conservation law corresponds to the duality symmetry? Infinitesimal duality rotation is performed on the electromagnetic potentials and the invariance of the Lagrangian (or action) is shown to lead to a conserved quantity: this conserved quantity is proportional to the difference in the number of right circularly polarized (RCP) and left circularly polarized (LCP) photons. Possible connection with Lipkin's zilch \cite{19} is also suggested. Note that the constant of motion for duality invariance given in \cite{25} also has correspondence with Lipkin's conserved zilch.

In an important work Zwanziger \cite{6} considers the transformation (3)-(4) for the Maxwell equations in vacuum in the presence of both electric and magnetic sources. He makes a distinction in terminology that for $\zeta = \pi /2$ the transformation is known as duality and for arbitrary $\zeta$ it is called chiral transformation. In the light of Calkin's conserved quantity related with helicity the term chiral appears appealing, and was used in \cite{8}. However here we adopt duality for arbitrary rotation angle $\zeta$. Zwanziger argues that for unitarily equivalent Hamiltonians under duality (chiral) transformation the photon state of momentum ${\bf k}$ and helicity $\lambda$ undergoes the transformation
\begin{equation}
 |{\bf k}, \lambda> ~ \rightarrow ~ U(\zeta)~|{\bf k}, \lambda> =e^{i\lambda\zeta}~|{\bf k}, \lambda> 
\end{equation}
A nice physical interpretation is given by him: the relative phase of left and right CP light or the absolute plane of polarization of linearly polarized light cannot be determined. The mathematical analysis of duality invariance is developed in a two-dimensional real vector space for the transverse radiation field variables ${\bf E}^r,~ {\bf H}^r$ and then introducing two vector potentials akin to those given  in Eq. (16). The generator $G$ in the unitary duality transformation
\begin{equation}
 U(\zeta) = e^{i\zeta G}
\end{equation}
is Hermitian, and determines the difference between the number of RCP and LCP photons. Regarding Dirac's charge quantization condition it is shown that duality invariant theory gives a new rule, namely the quantization of the chiral combination of electric and magnetic charges $(e_m g_n -g_m e_n)$.

A formal complex vector representation ${\bf \Psi} = {\bf E} + i {\bf B}$ renders the curl Maxwell equations (6) and (7) in a suggestive Schroedinger-like form \cite{10}
\begin{equation}
 {\bf S}.{\bf \nabla \Psi}=\frac{\partial {\bf \Psi}}{\partial t}
\end{equation}
where $3\times3$ matrices ${\bf S}$ are defined as
\begin{equation}
 {(S_i)}_{jk} = i \epsilon _{ijk}
\end{equation}
with $\epsilon_{ijk}$ the Levi-Civita tensor in three dimension. Multiplying by $i\hbar$ on both sides of Eq. (19) and making the identification ${\bf p} = -i\hbar {\bf \nabla}$ it follows that the Hamiltonian is $H =-{\bf S}.{\bf p}$. The duality rotation (3) and (4) assumes the form
\begin{equation}
 {\bf \Psi} ~ \rightarrow ~ e^{-i\zeta} {\bf \Psi}
\end{equation}
Note that the divergence equations (5) and (8) combine to give
\begin{equation}
 {\bf \nabla}.{\bf \Psi} =0
\end{equation}
Assuming that initially both electric and magnetic fields have zero divergence one could treat Eq. (22) as a subsidiary condition. Drawing analogy of photon equation (19) with Weyl equation for massless neutrino, and postulating two elementary fields in two dimensional real vector space the chiral invariance or duality invariance was used in \cite{26} to speculate on the nature of monopole and composite photon. An interesting review by Kobe \cite{27} shows that relativistic Schroedinger-like or rather more appropriately Dirac's relativistic spinor-like equation for photon is a fascinating subject, and has a long history. Preceding discussion indicates the profound physical significance of duality symmetry in electromagnetism. Regarding the difference in implementing duality in Maxwell equations and action integral two major issues elaborated below seem to be crucial.

${\bf A.}~$ Fundamental dynamical field variables

The action (9) in manifestly covariant form can be written as
\begin{equation}
 I_c = -\frac{1}{4} \int F^{\mu\nu} F_{\mu\nu}~ d^4x
\end{equation}
and recall that it is invariant under the gauge transformation
\begin{equation}
 A_\mu  \rightarrow A_\mu + \partial_\mu \chi
\end{equation}
In the standard theory the Lagrangian density in the Lorentz scalar action integral (23) has a functional dependence on independent field variable $A_\mu$ and its derivatives $\partial_\mu A_\nu$. Variational principle in the usual way for infinitesimal variations $\delta A_\mu$ and $\delta(\partial_\mu A_\nu)$ gives rise to the Euler-Lagrange equation of motion
\begin{equation}
 \partial_\mu F^{\mu\nu} = 0
\end{equation}
Eq. (25) is a Lorentz covariant representation of half of the full set of Maxwell equations, i. e. only Eqs. (5) and (6). The obvious and well known fact is that $A_\mu$ does not appear in Maxwell equations: Eqs. (5)-(8) or Eqs. (19) and (22) or Eq. (25). However the definition of $F_{\mu\nu}$ could be viewed as a re-statement of one of the pairs of Maxwell equations (7) and (8). To see it in a more explicit way, using elementary vector calculus Eq. (8) implies that
\begin{equation}
 {\bf B} = {\bf \nabla} \times {\bf A}
\end{equation}
and substituting (26) in (7) we have
\begin{equation}
 {\bf E} = - {\bf \nabla} \phi - \frac{\partial {\bf A}}{\partial t}
\end{equation}
Expressions (26) and (27) define the electromagnetic field tensor $F_{\mu\nu}$. It is evident that variation of action (23) does not give the full set of Maxwell equations as the equations of motion. For the sake of completeness we write the remaining pair also in compact covariant form
\begin{equation}
 \partial_\mu ~^* F^{\mu\nu} =0
\end{equation}
where the dual tensor $^*F^{\mu\nu} =\frac{1}{2} \epsilon^{\mu\nu\rho\sigma}~F_{\rho\sigma}$. All of this is a textbook matter but given here for the added emphasis on some aspects: when duality invariance or lack of it is discussed the distinction between the experimental laws abstracted in the form of complete Maxwell equations, equations of motion derived from variational principle and functional form of action has to be kept in mind. It is true that the currently accepted view is, what, for example, Witten underlines \cite{2} that four vector potential has fundamental importance in 20-th century physics. As we have noted above $A_\mu$ is a basic dynamical variable in the action for electromagnetism. However from the experimental point of view in classical electrodynamics $A_\mu$ is superfluous or at best an auxiliary convenient mathematical tool. It may be mentioned that though observed Aharonov-Bohm effect is a manifestation of vector potential, treating it as a typical quantum effect the debate on the physical reality of electromagnetic potentials continues unabated. It seems the origin of the failure of local duality invariant theory lies at the level of ambiguity in implementing global duality rotation itself in the action, and thus the review by Saa \cite{28} in response to \cite{7, 13} though interesting seems to be of limited scope. Could one dispense with $A_\mu$ completely and construct action purely with electromagnetic field as independent dynamical variable? If one could do it the aforementioned problems would not arise. Remarkably Sudbery \cite{9} achieves this goal but then paying a price since the action is not conventional Lorentz scalar but a pseudo-four vector. We present new results based on this action in the next section.

The problem persists also in the Hamiltonian formulation. See Dirac's quite elegant analysis on the role of constraints in Hamiltonian field theory \cite{29}, a good introductory treatment in Sec. 24, Ch. III of \cite{20} and with emphasis on gauge theories in Ch. 7 of \cite{30}. In the present context the language of constraints used by the authors in \cite{14, 25} in both Hamiltonian and Lagrangian forms helps in formal reconciliation of duality invariance in the action: Eq. (5) above is re-named as Gauss constraint, Eq. (26) as algebraic constraint, and Eq. (27) as an identity not a field equation. Another point made by the authors is that no gauge condition is imposed; only the transversality of fields is used to describe the system by a reduced set of field variables. Note that Zwanziger \cite{6} precisely deals with such a reduced set of radiation fields.

We argue that possibly this approach succeeds as a result of an ambiguity in the construction of action for the source-free case: it is a matter of convention that the field tensor is defined by (26) and (27) that essentially embody Maxwell equations (7) and (8); one could alternatively define electric and magnetic field vectors (with a sign change) by (26) and (27) respectively that would represent the first pair of Maxwell equations (5) and (6) and the variation in action would give rise to the equations of motion (7) and (8). If we insist on the physical origin of Maxwell equations none of the choices could be treated as mere constraints. It is significant that the standard definition of the field tensor retains the distinction between the pair of equations that emanates in the presence of sources.

${\bf B.}~$ Time and relativistic invariance

Corson in Sec. 19 Ch. III \cite{20} presents a conceptual analysis of the action principle in field theories. Of special interest here is the role of time while performing the variations since reference to particular Lorentz frames spoils relativistic invariance. He notes that instead of the volume between surfaces of constant time an invariant concept of space-like surfaces could be used. In a very lucid discussion Dirac \cite{29} points out the special role that time plays singling out a specific observer for developing the theory of Hamiltonian dynamics. Now even after assuming a preferred time coordinate the canonically conjugate momentum variable for Maxwell action (23)
\begin{equation}
 \pi _\mu = \frac{\partial L}{\partial \dot{A} _\mu}
\end{equation}
poses a serious problem as its time component vanishes. As a consequence there arises inconsistency with the fundamental Poisson bracket relation. The procedure to obtain the final form of physically acceptable Hamiltonian \cite{30} involves intermediate gauge transformations as one takes the system from one point of time to another. Deser and Teitelboim implement duality in terms of time-local variations: the change in action under duality rotation is obtained in the form of total time derivatives; see Eqs. (2.4) and (2.13) in their paper \cite{25}. The implicit role of using superfluous gauge variable could be seen there to arrive at a set of reduced dynamical variables: gauge invariant transverse fields/vector potential. Ramond \cite{30} gives examples of Coulomb and Arnowitt-Fickler gauges and makes a remark to the effect that one has to be careful about the difference between a non-dynamical variable and a genuine gauge condition. Is there some non-obvious role of gauge condition in implementing duality transformation on the action? We have not been able to arrive at a definite answer and suggest that this question and the role of time needs further examination to avoid the likely source of confusion in the formalism.

\section{\bf Local duality in Sudbery's formalism}

Sudbery's unconventional formalism \cite{9} is motivated by a new variational principle in which electric and magnetic field vectors are independent dynamical variables not the standard electromagnetic four vector potential; it was proposed by Anderson and Arthurs \cite{15} and independently by Rosen \cite{16}. Novel and intriguing features of Sudbery's action are that it is a pseudo-four vector not a conventional Lorentz scalar, and Noether's theorem leads to the conserved energy-momentum tensor as a consequence of duality invariance whereas the invariance under spacetime translations leads to a conserved third rank tensor that is related to Lipkin's tensor \cite{19}. My interest in this formalism \cite{8} arose due to the interesting role of duality in it, however it must be made clear from the outset that duality invariance was not the motivation or the focus of attention in \cite{9, 15, 16} as it is in the present paper.

Let us begin with the action proposed in \cite{15, 16}
\begin{equation}
 I_{AAR} = \int ( {\bf B}.\frac{\partial {\bf E}}{\partial t} - {\bf E}.\frac{\partial {\bf B}}{\partial t} -{\bf E}.({\bf \nabla} \times
{\bf E})-{\bf B}.({\bf \nabla} \times {\bf B})+ 2 {\bf J}. {\bf B}) ~ d^4 x
\end{equation}
which is a pseudo-scalar, and the variations in ${\bf E}$ and ${\bf B}$ in the usual way lead to the following Euler-Lagrange equations
\begin{equation}
{\bf \nabla} \times {\bf E} = - \frac{\partial {\bf B}}{\partial t} 
\end{equation}
\begin{equation}
{\bf \nabla} \times {\bf B} = {\bf J} + \frac{\partial {\bf E}}{\partial t}  
\end{equation}
In the preceding section the ambiguity in the Mawxwell action was noted that either of the pairs (5) and (6) or (7) and (8) could correspond to the Euler-Lagrange equations depending on the definition of the field tensor; here interestingly the pair (31) and (32) represents the curl equations (6) and (7) (setting the current density ${\bf J}=0$). Sudbery notes two deficiencies of (30): first it is not in a Lorentz covariant form, and second that only partial set of Maxwell equations are obtained from it. Regarding the second it may be recalled that only half of the Maxwell equations follow as Euler-Lagrange equations from the standard action (23), however the definition of the field tensor implicitly contains the remaining pair. Since electric and magnetic field vectors in (30) are treated as fundamental field variables one cannot get the divergence equations in the AAR formulation.

The Lagrangian density for ${\bf J} =0$ in the action (30) can be written in an elegant form using the complex vector representation for the fields ${\bf \Psi}$ similar to the Good's formalism \cite{10} discussed in Sec. III. The new form is given by
\begin{equation}
L_{AAR} = \frac{1}{2i} ({\bf \Psi} \frac{\partial {\bf \Psi}^*}{\partial t} -{\bf \Psi^*} \frac{\partial {\bf \Psi}}{\partial t}-{\bf\Psi ~S.\nabla\Psi}^* - {\bf \Psi}^* ~ {\bf S.\nabla \Psi})
\end{equation}
It is easy to verify that (33) is invariant under the duality rotation (21). Though we do not pursue it here this form of Lagrangian seems interesting to seek its generalization in analogy with that of Dirac's relativistic spinor field.

Sudbery makes a radical proposal in which Lagrangian is a pseudo-four vector and the action is defined to be
\begin{equation}
 I^s_\sigma = \int L^s_\sigma ~ d^4 x
\end{equation}
\begin{equation}
 L^s_\sigma =^*F^{\mu\nu}\partial_\nu F_{\mu\sigma} - F^{\mu\nu} \partial_\nu ~ ^*F_{\mu\sigma} - 2~ ^* F_{\sigma \mu} J^\mu
\end{equation}
Using the variational principle the complete set of Maxwell equations with sources are obtained as Euler-Lagrange equations. In the source free case the action is invariant under the duality transformation
\begin{equation}
 F_{\mu\nu} = F_{\mu\nu} cos \zeta + ^*F_{\mu\nu} sin \zeta
\end{equation}
\begin{equation}
 ^*F_{\mu\nu} = - F_{\mu\nu} sin \zeta + ^*F_{\mu\nu} cos \zeta
\end{equation} 
and using Noether's theorem one gets the conservation law for symmetric energy-momentum tensor $T_{\mu\nu}$. Another intriguing result is that the invariance under spacetime translations leads to conserved third rank tensor related with Lipkin's tensor \cite{19}. It is noteworthy that duality invariance leads to the conservation of symmetric energy-momentum tensor in contrast to nonsymmetric and gauge noninvariant canonical tensor $E_{\mu\nu}$ obtained in the standard formulation as a consequence of spacetime translation symmetry of the action (23). Recalling that the term added to $E_{\mu\nu}$ to obtain $T_{\mu\nu}$ has an interpretation corresponding to spin energy \cite{20} it seems duality rotation has a significant role in the polarization and angular momentum of light.

Sudbery's Lagrangian can be further generalized to incorporate local duality invariance \cite{8} in an unambiguous manner. The new Lagrangian density in the absence of sources is given by
\begin{equation}
L_\sigma = L^s_\sigma - g(F^{\mu\nu} F_{\mu\sigma} + ^*F^{\mu\nu}~ ^*F_{\mu\sigma}) W_\nu
\end{equation}
It is straightforward to verify that (38) is invariant under the transformations (36) and (37) with $\zeta$ being a function of spacetime provided the pseudo-four vector $W_\nu$ transforms as $W_\nu \rightarrow W_\nu + g^{-1} \partial_\nu \zeta$ where $g$ is a coupling constant. The variational procedure for action integral (34) with the integrand (38) in it gives rise to the field equations
\begin{equation}
\partial_\mu F^{\mu\nu} = g~ W_\mu ~^*F^{\mu\nu}
\end{equation}
\begin{equation}
\partial_\mu ~ ^*F^{\mu\nu} = - g~ W_\mu~ F^{\mu\nu}
\end{equation}
Equations (39) and (40) written in terms of electric and magnetic field vectors make physical content more transparent
\begin{equation}
 {\bf \nabla.E} = -g ~{\bf W.B}
\end{equation}
\begin{equation}
 {\bf \nabla.B} = g~ {\bf W.E}
\end{equation}
\begin{equation}
{\bf \nabla} \times {\bf E} + \frac {\partial {\bf B}}{\partial t} + g {\bf W} \times {\bf B} +g W_0 {\bf E} =0 
\end{equation}
\begin{equation}
 {\bf \nabla} \times {\bf B} - \frac {\partial {\bf E}}{\partial t} - g {\bf W} \times {\bf E} +g W_0 {\bf B} =0
\end{equation}
A nice property of Sudbery's Lagrangian is that it can be further generalized to include electric charge and current densities $J^\mu _e$ as well as magnetic charge and current densities $J^\mu _m$.  Notice that the Lagrangian density $L_{AAR}$ in (30) could be modified adding a term $-2 {\bf J}_m .{\bf E}$ that changes (31) to
\begin{equation}
 -{\bf \nabla} \times {\bf E} =  \frac{\partial {\bf B}}{\partial t} +{\bf J}_m
\end{equation}
Proposed new generalization of Lagrangian density (38) is
\begin{equation}
 L^N_\sigma = L_\sigma -2 ~ ^*F_{\sigma\mu} J^\mu _e + 2 F_{\sigma\mu} J^\mu_m
\end{equation}
It can be verified that under the local duality rotation (36) and (37) and simultaneous duality transformations of current densities
\begin{equation}
 J^\mu_e ~ \rightarrow ~ J^\mu_e cos \zeta + J^\mu_m sin \zeta
\end{equation}
\begin{equation}
 J^\mu_m ~ \rightarrow ~ -J^\mu_e sin \zeta + J^\mu_m cos \zeta
\end{equation}
the new action
\begin{equation}
 I^N_\sigma = \int L^N_\sigma ~ d^4 x
\end{equation}
is invariant. The Euler-Lagrange equations of motion derived from the action (49) using the variational principle \cite{9} are obtained to be the generalization of Eqs. (39) and (40)
\begin{equation}
 \partial_\mu F^{\mu\nu} = g~ W_\mu ~^*F^{\mu\nu}+ J^\nu_e
\end{equation}
\begin{equation}
 \partial_\mu ^*F^{\mu\nu} = - g~ W_\mu~ F^{\mu\nu}+ J^\nu_m
\end{equation}
The vector transcription of Eqs. (50) and (51) in terms of electric and magnetic field vectors is given by
\begin{equation}
{\bf \nabla.E} = -g ~{\bf W.B} + \rho_e
\end{equation}
\begin{equation}
 {\bf \nabla.B} = g~ {\bf W.E} +\rho_m
\end{equation}
\begin{equation}
{\bf \nabla} \times {\bf E} + \frac {\partial {\bf B}}{\partial t} + g {\bf W} \times {\bf B} +g W_0 {\bf E} =- {\bf J}_m
\end{equation}
\begin{equation}
 {\bf \nabla} \times {\bf B} - \frac {\partial {\bf E}}{\partial t} - g {\bf W} \times {\bf E} +g W_0 {\bf B} = {\bf J}_e
\end{equation}

The pseudo- four vector field $W_\mu$ plays the role of a duality gauge field; could it be promoted to a genuine dynamical field? Any added kinetic term for this field must be invariant under the gauge transformation, and therefore it has to be similar in form to the electromagnetic field tensor. The suggested additional Lagrangian density is
\begin{equation}
 L^g_\sigma = -\frac {1}{4} W^{\mu\nu} W_{\mu\nu} C_\sigma
\end{equation}
where
\begin{equation}
 W_{\mu\nu} = \partial_\mu W_\nu - \partial_\nu W_\mu
\end{equation}
Here $C_\sigma$ is an arbitrary Lorentz four-vector field; it is not a constant vector. The field equations (50) and (51) remain unchanged while infinitesimal variation $\delta W_\mu$ in the action gives rise to the following equation
\begin{equation}
 \partial_\nu (C_\sigma W^{\rho\nu}) = - g ( F^{\mu\rho} F_{\mu\sigma} + ^*F^{\mu\rho} ~ ^*F_{\mu\sigma})
\end{equation}
It is instructive to write the time components of (58) for the assumed gauge condition
\begin{equation}
 \partial_\mu W^\mu = 0
\end{equation}
\begin{equation}
 -C_0 \partial^\mu \partial_\mu W_0 + {\bf \nabla} C_0.({\bf \nabla} W_0 +\frac {\partial {\bf W}}{\partial t}) = g (E^2 + B^2)
\end{equation}
\begin{equation}
 -C_0\partial^\mu \partial_\mu {\bf W} - \frac {\partial C_0}{\partial t}({\bf \nabla} W_0 +\frac {\partial {\bf W}}{\partial t})= 2g({\bf E} \times {\bf B})
\end{equation}
In a special case assuming $C_0$ is constant Eqs. (60) and (61) formally resemble Wilson's equations for the gravitational vector potential \cite{31}; since the source terms are energy-momentum densities Wilson suggested a vector potential theory for gravitation. However such a theory disagrees with experiments, for example it cannot account for the observed perihelion advance of the planet Mercury. Further $W_\mu$ is a pseudo-vactor and space reflection symmetry being a good symmetry in gravitation we cannot identify it with gravitational potential. Could it be weak gauge boson? Parity violation in weak interactions and the existence of neutral weak gauge boson $Z$ would tempt one to consider this possibility, however $Z$ is a massive particle. Could $C_0$ play some role in giving mass to $W_\mu$? We do not have answer to these questions at present. However there are interesting possibilities to apply the present theory to physical phenomena discussed in the next section.

\section{\bf  Physical implications}

Electric-magnetic local duality could be implemented in a neat form and in an unambigous manner: it had been proved for source-free case in \cite{8} using Sudbery's pseudo-four vector Lagrangian density \cite{9} and was demonstrated in the Maxwell field equations in a Lorentz covariant form recently \cite{17}. In the present paper we have generalized Sudbery's action to local duality invariant form in the presence of electric and magnetic sources and showed that the Euler-Lagrange equations obtained from the new action are consistent with the generalized Maxwell equations proposed in \cite{17}. The most uncomfortable part of the present theory seems to be the use of a pseudo- four vector action rather than the traditional scalar action. It may be emphasized that relativistic invariance is maintained  in this approach. On the positive note, the equations of motion derived from Sudbery's action using the variational principle represent the complete set of the experimentally established Maxwell field equations thus providing a posteriori justification for it. Moreover this theory offers an alternative approach in which the electromagnetic fields are fundamental dynamical variables in the action. Could there be a more appealing conceptual justification to treat action as a vector? We do not know for sure; a plausible argument could be made remembering first that physical dimension of action and angular momentum vector are identical, and second that energy traditionally considered scalar, in the relativistic framework becomes a time component of energy-momentum four vector. Perhaps we have a hint here that deserves serious attention and search for concrete examples.

As we have seen Sudbery's theory is not a U(1) gauge theory and its quantization has not been attempted; in fact it is not known if quantization would succeed in this case. We have delineated subtle technical questions in Sec. III regarding the global duality symmetry in the standard action (23). It could be stated straightforwardly that in the U(1) gauge framework local duality cannot be implemented in the action formulation in agreement with the conclusion arrived at in \cite{7, 13}. The literature on duality symmetry in superstrings and supergravity \cite{1, 2} and elegant elucidation of SL(2, Z) symmetry in condensed matter and high energy physics by Fradkin and Kivelson \cite{4} show that the application of the new local duality invariant theory in them is not clear. Bunster and Henneaux \cite{7} in Sec. IV of their paper point out that gauging electric-magnetic duality in supergravity is not possible; since vector potentials are dynamical fields in this context it is not immediately obvious if the present ideas could be useful.

Aforementioned drawbacks do not imply that the present theory is just a vacuous intellectual curiosity rather it holds promise for new developments in mathematical physics and has important physical implications. First, consider the Lagrangian density (33) for complex vector field representation of electromagnetism - it could be generalized to implement local duality and possibly could be quantized following the approach given by Kobe \cite{27}. From the historical point of view we must mention that Wilson \cite{31} independently of Anderson and Arthurs \cite{15} and Rosen \cite{16} discovered the Lagrangian density (33) and also derived the local duality invariant Lagrangian and the Euler-Lagrange equations from it. We have pointed out that Noether's theorem applied to Sudbery's action \cite{9} leads to new conserved quantities. There is no general theory where unconventional proposition of four vector action, symmetries and conservation laws have been developed in the spirit of Noether's theorem; such a work would open new avenues in mathematical physics.

The most significant utility of the new theory would be in the macroscopic classical phenomena; below we discuss two areas of current interest.

{\bf The nature of polarized light}

Enigma of photon and its precise relationship with the properties of light, in particular, polarization continues to inspire new experiments in (quantum !) optics and foundational discourse on the physical reality of photon. The present discussion is limited to certain aspects where the new generalized theory could be applicable to advance our understanding on them. First we mention vacuum birefringence predicted by Heisenberg- Euler effective Lagrangian for quantum electrodynamics. Though it has been known since long it acquired renewed interest recently motivated by experiments and searches for elusive axions. Phenomenologically one can incorporate this effect in terms of polarization tensors and modified constitutive relations in Maxwell equations, see \cite{32} for a concise discussion and references.

Birefringence in material media is another area where intriguing features on polarized light propagation have emerged. Chiral media represent an important class of materials: a chiral medium has mirror asymmetry and RCP and LCP light interact differently in such a medium. Optical rotation in a natural optically active medium can be understood in terms of circular birefringence. To probe the light - media interaction one could study reflection of light from a plane boundary of chiral-achiral interface, reflection from an achiral-chiral interface and lastly reflection from achiral-achiral interface. A nice comprehensive discussion on these problems can be found in \cite{33}; the authors Silverman and Badoz bring out the importance of great experimental ingenuity in such investigations, moreover there are controversial issues pertaining to theoretical interpretation. What are the correct constitutive relations? What are the proper boundary conditions? Also related with them is the question of conserved quantities. In an excellent monograph \cite{34} Post presents a profound analysis on the general covariance of Maxwell equations and moving a step ahead of general covariance underlines the significance of their (i. e. of Maxwell equations) natural invariance (metric-independence). The role of symmetry on constructing constitutive relations becomes quite transparent. In Chapter 6 of the electron monograph \cite{35} different mathematical representations of Maxwell equations and modifications are reviewed; Imbert's shift and Jones-Richards experiment are discussed in \cite{8}. Regarding initial boundary value problem we refer to a notable contribution in \cite{36}.

Duality has fundamental significance for chiral phenomena and polarized light as discussed in Sec. III, also see \cite{8}. In the source-free vacuum case local duality invariant Maxwell equations (41) - (44) can be re-written in terms of macroscopic fields ${\bf D}$ and ${\bf H}$ defining new constitutive relations; these could be useful for a class of magneto-electric materials, see for example the references in \cite{28}. In the presence of material medium the most general representation of the constitutive relations constrained by duality invariance can be obtained from the generalized Lagrangian density (46) introducing the polarization currents in addition to source currents following the approach of \cite{17}
\begin{equation}
 J^\mu_P = \partial_\nu ~ P^{\nu\mu}
\end{equation}
\begin{equation}
 ^*J^\mu_P = \partial_\nu ~ ^*P^{\nu\mu}
\end{equation}
Here $^*P^{\mu\nu}$ is dual to antisymmetric polarization tensor $P^{\mu\nu}$ and $P^{0i}$ and $P^{ij}$ correspond to electric ${\bf P}$ and magnetic ${\bf M}$ polarization vectors respectively. The Euler-Lagrange equations (50) and (51) get modified to
\begin{equation}
 \partial_\mu F^{\mu\nu} = g~ W_\mu ~^*F^{\mu\nu}+ J^\nu_e +J^\nu_P
\end{equation}
\begin{equation}
  \partial_\mu ~^*F^{\mu\nu} = - g~ W_\mu~ F^{\mu\nu}+ J^\nu_m +^*J^\nu_P 
\end{equation}

In the absence of sources $J^\mu_e$ and $J^\mu_m$  and vanishing duality gauge potential $W_\mu$ it becomes straightforward to write the generalized Maxwell equations in terms of the tensor $G^{\mu\nu} = F^{\mu\nu} - P^{\mu\nu}$ that embodies the duality symmetric constitutive relations. Thus generalized Maxwell field equations (50) - (51) and (64) - (65) give a unified picture for the propagation of electromagnetic waves in different kinds of media including birefringent vacuum.

Topological properties of light have been a subject of intense research activity for past more than two decades, specially those associated with geometric phases and vortices. Discovery of Berry phase in nonrelativistic quantum mechanics in 1984 and subsequent recognition \cite{37} that Pancharatnam in 1956 \cite{12} had observed a nontrivial geometric phase in polarization optics stimulated enormous work in this field. A polarized light wave propagating along a fixed direction in space subjected to undergo a polarization cycle in the polarization state space acquires a geometric phase named after Pancharatnam. Poincare sphere is a nice geometric representation of polarization state of light, and the geometric phase equals half of the solid angle subtended by the the closed cycle. Mathematically Pancharatnam phase can be understood as a consequence of the change in the direction of a vector parallel transported on the surface of the sphere. Is there any physical origin of this effect?

Singular optics or optical vortices have roots in the topological defects in continuous fields: wavefront dislocations or phase singularities for scalar waves \cite{38} and disclinations or polarization singularities for vector waves \cite{39}. In a recent article \cite{40} the experimental evidence to relate angular momentum (AM), topological defects, and geometric phases in optics has been critically reviewed. Another geometric phase, namely spin redirection phase results when the cyclic change is made in the momentum state space of light wave that we have termed as Reetov-Vladimirskii-Chiao-Wu (RVCW) phase \cite{40}. Interestingly in RVCW phase there is a sort of geometric circular birefringence. In 1992 we put forward AM holonomy conjecture as a common physical origin for geometric phases in optics. Spin angular momentum (SAM) is related with polarization of light and since there is a sequence of polarization changes in Pancharatnam phase it is natural to expect SAM exchange in this process. Cycles in momentum space would entail changes in orbital angular momentum (OAM) of light indicating the role of OAM exchange in RVCW phase. AM holonomy conjecture is founded on a re-interpretation of a well known result in standard electrodynamics that emanates from constructing gauge invariant and Lorentz covariant conserved quantities from Noether conserved quantities. A very good discussion can be found in Corson's monograph \cite{20} and for a shorter one see \cite{23} ; curiously Noether's theorem is not mentioned in either of the books.

Consider the Maxwell action (23) then Noether's theorem shows that infinitesimal spacetime translational invariance leads to conserved canonical energy-momentum tensor $E^{\mu\nu}$ and infinitesimal Lorentz transformation invariance gives a conserved third rank AM tensor $M^{\alpha\mu\nu}$. One can split the AM tensor into SAM and OAM parts
\begin{equation}
 M^{\alpha\mu\nu} = - (E^{\alpha\mu} x^\nu -E^{\alpha\nu} x^\mu) - (F^{\alpha\mu} A^\nu - F^{\alpha\nu}A^\mu) = L^{\alpha\mu\nu} +S^{\alpha\mu\nu}
\end{equation}
Note that $E^{\mu\nu}$ and $M^{\alpha\mu\nu}$ are natural Noether conserved quantities for the action (23); unfortunately these are not gauge invariant, energy-momentum tensor is not symmetric and the separation (66) is not Lorentz invariant. There is a well known prescription to construct symmetric and gauge invariant energy-momentum tensor $T^{\mu\nu}$ and using it to define a third rank AM tensor $J^{\alpha\mu\nu}$. These new quantities are believed to be physically observable ones. The two AM tensors are related as
\begin{equation}
 M^{\alpha\mu\nu} = J^{\alpha\mu\nu} - \partial_\lambda (F^{\lambda\alpha} (A^\mu x^\nu - A^\nu x^\mu))
\end{equation}
In the usual way one argues that the integral of the total divergence term in (67) could be made to vanish having no physical significance. In contrast to this the AM holonomy conjecture is founded on making an analogy with Aharonov-Bohm effect: this term could make a topological contribution in nontrivial situations and this manifests as a geometric phase.

Could one give a firm foundation for this conjecture? Advancement in duality invariant theory presented in Sec. IV offers a possible approach to achieve this goal. Qualitative arguments \cite{8} based on the role of duality symmetry in polarized light phenomena \cite{5, 6} having bearing on geometric phases, and a simple duality gauge theory \cite{11} generalizing Good's formulation \cite{10} to explain Pancharatnam phase establishes a connection between duality symmetry and geometric phases in optics. Duality symmetry in Sudbery's theory \cite{9} surprisingly leads to Noether conserved quantity $T^{\mu\nu}$. What is the physics behind it? This question led us to develop local duality generalization of Sudbery's action in \cite{8}. It is plausible to argue that duality gauge potential mediates AM transfer as was first proposed in \cite{11}. It is further supported considering the Lagrangian density (38) and adding a topological term for $W^\mu$ similar to (13) that gives a topological effect. Such a form is suggested by Sudbery in a private communication. To prove AM holonomy conjecture the total derivative term in (67) may have to be related with duality gauge potential and it may become necessary to analyze technical aspects in dealing with two kinds of gauge potentials $A^\mu$ and $W^\mu$. We do not know how to do it at present, however in the light of an existing gauge theory for Pancharatnam phase \cite{11} and a recent work \cite{41} that gives a gauge theoretic explanation for geometric phases in astigmatic optical modes this tentative idea deserves serious attention.

There is another intriguing result in Sudbery's theory, namely the appearance of a third rank tensor as a Noether conserved quantity for spacetime translational invariance that is related to Lipkin's tensor $Z^{\alpha\mu\nu}$. Lipkin in section 8 of his paper \cite{19} calculates zilch density tensor $Z^{\alpha\mu\nu}$ for a plane monochromatic electromagnetic wave and finds that it depends on the polarization state of the wave. For a linearly polarized wave there is no transport of zilch, but there is a flow of zilch at the rate proportional to the frequency of the wave and oppositely directed for LCP and RCP waves. It would be interesting to calculate $Z^{\alpha\mu\nu}$ for OAM bearing light beams \cite{42} and those with radial and azimuthal polarization singularities \cite{39, 40}. In case one discovers a classification scheme based on this, this classification along with SAM and OAM characterization may throw light on recent apparently counter-intuitive phenomena reported in the literature, see \cite{40} for references.

{\bf Topological insulators}

The subject of topological insulators is a fast developing and exciting field in condensed matter. Qi and Zhang \cite{43} begin their expository article on this recalling quantum Hall effect as a first distinctly topological system and interestingly SL(2, Z) symmetry seems to have been applied first for this system \cite{4}. Recently Karch \cite{44} presented a duality perspective to topological insulators, and though no new result is given the idea of duality in topological insulators is an attractive one. Noninteracting topological band theory and topological field theory are believed to describe satisfactorily topological insulators \cite{43}. For a three dimensional topological insulator an effective action that resembles (12) has been of great value in understanding its electromagnetism. Could one envisage it as a duality rotation induced effect? Focusing on the physics of topological insulators in \cite{17} we have derived generalized local duality invariant Maxwell equations and shown that topological magneto-electric effect and axion-like electrodynamics could be obtained from this. In the present paper the main theme is that of duality symmetry, however consistency of the Euler-Lagrange equations of motion derived here with the generalized Maxwell equations given in \cite{17} suggests the role of duality invariant theory for topological insulators at a fundamental level in which duality gauge potential $W^\mu$ rather than the elusive axion has physical significance.

On the experimental side optical studies of topological insulators are very interesting \cite{45}. Recalling our discussion on chiral media interfaces investigated using polarized light \cite{33} and the interesting features of optical vortices noted above it would be desirable to probe the surfaces of topological insulators using such beams.

In conclusion, we have presented a fresh outlook on global duality symmetry and developed a local duality invariant theory of electromagnetism in the most generalized form, and discussed the mathematical and physical implications of this theory.

{\bf APPENDIX}

Following response from known/unknown physicists we address briefly a question regarding the vector field $C^\mu$. The role of the field $C_\mu$ in the additional Lagrangian density (56) has to be clearly explained keeping in mind the subtle aspects on relativistic invariance and superfluous gauge variables briefly presented in point B of Sec. II following the treatments given in \cite{20, 29, 30}. The gauge potential $W_\mu$ is natural in a local duality gauge theory, however unlike it the field $C_\mu$ is a nondynamical auxiliary field. It is similar to spinless fields $F$ and $G$ introduced in the Wess-Zumino supersymmetric action, see Section 1.8 of \cite{30}. Note that we are not discussing supersymmetry here, the point is just to draw attention to the existence of a nondynamical field variable. Auxiliary fields having no dynamics or kinetic energy terms are also well known in implementing BRST symmetry in gauge field theories. Thus we do not need extension of expression (56) for making $C_\mu$ a dynamical field variable. It is important to realize that Eqs. (60) and (61) are also consistent with the usual gauge theoretic prescription. Global gauge symmetry leads to a conserved Noether current and for local gauge invariance one obtains the conserved current as the source of the gauge field. In the present case, global duality invariance of Sudbery's action gives rise to energy-momentum as conserved Noether current and this current appears as the source term in Eqs. (60) and (61). More delicate is the issue of relativistic invariance at the level of the principle in the action, however the equations of motion derived from it, (50), (51) and (58) are Lorentz covariant. Thus within this limitation the generalized local duality invariant theory developed here is complete.

One could, of course, raise an interesting question: Could we dispense with the necessity of an auxiliary field altogether? We provide the answer in affirmative and present an alternative. Instead of $L^g_\sigma$ given by expression (56) let us assume a term $L^{ga}_\sigma$ proportional to $\epsilon_{\mu\nu\rho\sigma} W^{\mu\nu} W^\rho$ for the additional Lagrangian density corresponding to the duality gauge field. Its similarity to the topological current (13) may be noted. This form was suggested by A. Sudbery in a private communication. It is important to note that the generalized Maxwell field equations (50) and (51) would remain unaltered ( there would be no auxiliary field $C^\mu$ ). Equation of motion for the duality gauge field $W_\mu$ would be different than Eq. (58), however none of the physical implications based on the generalized Maxwell equations discussed in the last section would change. Though the auxiliary field $C^\mu$ gives somewhat more freedom in the equation of motion for the duality gauge field $W^\mu$ the Euler-Lagrange equation derived from the action with the alternative term $L^{ga}_\sigma$ is given below for the sake of completeness
\begin{equation}
2\epsilon_{\alpha\beta\mu\sigma} \partial ^\alpha ~W^\beta ~+~\epsilon_{\alpha\beta\mu\sigma} W^{\alpha\beta} =~g(F_\mu^\rho F_{\rho\sigma} +~^*F_\mu^\rho F_{\rho\sigma})
\end{equation}
The time component of the above equation gives an interesting result: the curl of the duality gauge field ${\bf W}$ is proportional to the Poynting vector.

A new result presented in the paper on topological insulators \cite{17} needs to be re-emphasized in the context of duality: one could treat constitutive relations as dual to source currents; Eqs. (13) and (14) would be dual to Eqs. (17) and (18) in \cite{17}. The vector transcription of generalized Maxwell equations, i. e. Eqs. (19) to (22) in \cite{17} may also be corrected with the ones given in the present paper; of course the results of \cite{17} would not change. 

Acknowledgements

I dedicate this paper to R. V. Jones - a great insightful experimentalist in optics. I thank Professor A. Sudbery for his interest in this work and Dr. A. Karch for pointing out reference \cite{4}, and Professor E. Fradkin for encouraging communications. Library facility at Banaras Hindu University, Varanasi is acknowledged.


\begin{thebibliography}{99}
\bibitem{1} M. J. Duff, Electric/magnetic duality and its stringy origins, arXiv: hep-th/9509106
\bibitem{2} E. Witten, Duality, spacetime and quantum mechanics, Phys Today, 50(5), 28-33 (1997)
\bibitem{3} A. Zee, Quantum Field Theory In A Nutshell (Princeton Univ. Press, 2003)
\bibitem{4} E. Fradkin and S. Kivelson, Modular invariance, self-duality and the phase transition between quantum Hall plateaus, Nucl. Phys. B 474(FS), 543-574 (1996) 
\bibitem{5} M. G. Calkin, An invariance property of the free electromagnetic field, Am. J. Phys. 33, 958-960 (1965)
\bibitem{6} D. Zwanziger, Quantum field theory of particles with both electric and magnetic charges, Phys. Rev. 176, 1489-1495 (1968)
\bibitem{7} C. Bunster and M. Henneaux, Can (electric-magnetic) duality be gauged?, Phys. Rev. D 83, 045031(7pp) (2011)
\bibitem{8} S. C. Tiwari, Polarized light and chiral invariance, Phys. Essays, 4, 212-216 (1991)
\bibitem{9} A. Sudbery, A vector Lagrangian for the electromagnetic field, J. Phys. A Math. Gen. 19, L33-L36 (1986)
\bibitem{10} R. H. Good, Jr. Particle aspect of the electromagnetic field equations, Phys. Rev. 105, 1914-1919 (1957)
\bibitem{11} S. C. Tiwari, On the Berry phase, Phys. Lett. A 149, 223-224 (1990)
\bibitem{12} S. Pancharatnam, Generalized theory of interference, and its applications, Proc. Indian Acad. Sciences XLIV (5) Sec. A, 247-262 (1956)
\bibitem{13} S. Deser, No local Maxwell duality invariance, Class. Quantum Grav. 28, 085009(4pp) (2011)
\bibitem{14} S. Deser, Off-shell electromagnetic duality invariance, J. Phys. A: Math. Gen. 15, 1053-1054 (1982)
\bibitem{15} N. Anderson and A. M. Arthurs, A variational principle for Maxwell's equations, Int. J. Electron. 45, 333-334 (1978)
\bibitem{16} J. Rosen, Redundency and superfluity for electromagnetic fields and potentials, Am. J. Phys. 48, 1071-1073 (1980)
\bibitem{17} S. C. Tiwari, Role of local duality invariance in axion electrodynamics of topological insulators, arXiv: 1109.0829
\bibitem{18} T. Fulton, F. Rohrlich and L. Witten, Conformal invariance in physics, Rev. Mod. Phys. 34, 442-457 (1962)
\bibitem{19} D. M. Lipkin, Existence of a new conservation law in electromagnetic theory, J. Math. Phys. 5, 696-700 (1964)
\bibitem{20} E. M. Corson, Introduction to Tensors, Spinors and Relativistic Wave Equations (Blackie and Son, 1953)
\bibitem{21} P. A. M. Dirac, The theory of magnetic poles, Phys. Rev. 74, 817-830 (1948)
\bibitem{22} S. C. Tiwari, Time asymmetry in classical electrodynamics, Phys. Lett. A 121, 156-158 (1987)
\bibitem{23} J. D. Jackson, Classical Electrodynamics (John Wiley, Second Edition, 1975)
\bibitem{24} P. K. Townsend, Unity from duality, Physics World 8, 41-46 (1995); see also, C. M. Hull and P. K. Townsend, Unity of superstring dualities, Nocl. Phys. B 438, 109-137 (1995)
\bibitem{25} S. Deser and C. Teitelboim, Duality transformations of Abelian and non-Abelian gauge fields,  Phys. Rev. D13, 1592-1597 (1976)
\bibitem{26} S. C. Tiwari, Neutrino, photon and monopole, Nat. Acad. Sci. Letters 3, 243-247 (1980)
\bibitem{27} D. H. Kobe, A relativistic Schroedinger-like equation for a photon and its second quantization, Found. Phys. 29, 1203-1231 (1999)
\bibitem{28} A. Saa, Local electromagnetic duality and gauge invariance, Class. Quantum Gravity 28, 127002 (2011)
\bibitem{29} P. A. M. Dirac, Lectures on Quantum Mechanics (Yeshiva University, N. Y. 1964)
\bibitem{30} P. Ramond, Field Theory: A Modern Primer (Addison-Wesley, Second Edition 1990)
\bibitem{31} W. Wilson, Internal gauge symmetries and the gravitational field, Gen. Rel. Grav. 12, 51-55 (1980)
\bibitem{32} S. L. Adler, Vacuum birefringence in a rotating magnetic field, J. Phys. A: Math. Theor. 40, F143-F152 (2007)
\bibitem{33} M. P. Silverman and J. Badoz, Light reflection from a naturally optically active birefringent medium, J. Opt. Soc. Am A 7, 1163-1173 (1990)
\bibitem{34} E. J. Post, Formal Structure of Electromagnetics (North Holland, 1962; Dover Edition, 1997)
\bibitem{35} S. C. Tiwari, Rebirth of the Electron: Electromagnetism An unorthodox new approach to fundamental problems in physics (Self-published, 1997; published at www.lulu.com in 2006)
\bibitem{36} P. Hillion and A. Lakhtakia, On an initial boundary value problem involving Beltrami- Moses fields in electromagnetic theory, Phil. Trans. Physical Sciences and Engineering, 344, no. 1671, 235-248 (1993)
\bibitem{37} M. V. Berry, The adiabatic phase and Pancharatnam phase for polarized light, J. Mod. Opt. 34, 1401-1407 (1987)
\bibitem{38} J. F. Nye and M. V. Berry, Dislocations in wave trains, Proc. R. Soc. Lond. A 336, 165-190 (1974)
\bibitem{39} J. F. Nye, Polarization effects in the diffraction of electromagnetic waves: the role of disclinations, Proc. R. Soc. Lond. A  387, 105-132 (1983)
\bibitem{40} S. C. Tiwari, Topological defects, geometric phases, and the angular momentum of light, Optik - Int. J. Light Electron. Optics, 120, 414-417 (2009)
\bibitem{41} S. J. M. Habraken and G. Nienhuis, Geometric phases in astigmatic optical modes of arbitrary order, J. Math. Phys. 51, 082702 (19pp) (2010)
\bibitem{42} L. Allen, M. J. Padgett and M. Babiker, The orbital angular momentum of light, Prog Opt. 39, 291-372 (1999)
\bibitem{43} X-L Qi and S-C Zhang, The quantum spin Hall effect and topological insulators, Phys. Today 63(1), 33-38 (2010)
\bibitem{44} A. Karch, Electric-magnetic duality and topological insulators, Phys. Rev. Lett. 103, 171601 (4pp) (2009)
\bibitem{45} D. Hsieh, J. W. McIver, D. H. Torchinsky, D. R. Gardner, Y. S. Lee and N. Gedik, Nonlinear optical probe of tunable surface electrons on a topological insulator, Phys. Rev. Lett 106, 057401(4pp) (2011) 
 
\end{thebibliography}
\end{document}